\documentclass[twocolumn,PLB,showpacs,superscriptaddress]{revtex4}%
\usepackage{amsfonts}
\usepackage{amsmath}
\usepackage{amssymb}
\usepackage{graphicx}%
\begin{document}
\title{Noncommutative information is revealed from the static detector outside the
black hole }
\author{Jingran Shi}
\affiliation{School of Mathematics and Physics, China University of Geosciences, Wuhan
430074, China}
\author{Yipeng Liu}
\affiliation{School of Mathematics and Physics, China University of Geosciences, Wuhan
430074, China}
\author{Baocheng Zhang}
\email{zhangbaocheng@cug.edu.cn}
\affiliation{School of Mathematics and Physics, China University of Geosciences, Wuhan
430074, China}

\begin{abstract}
We investigate the transition behavior of the two-level atom as the
Unruh-DeWitt detector outside a noncommutative black hole. When the mass of
the black hole is small enough, the difference between the commutative and
noncommutative black hole can be distinguished. In particular, an evident
fluctuation appearing at a far distance from the horizon by calculating the
quantum Fisher information of the transition rate with regard to the local
Hawking temperature provides a novel and interesting result about the
information extraction of the noncommutativity for a small-mass black hole.

\end{abstract}
\maketitle

\section{Introduction}

The Unruh-Dewitt (UDW) detector \cite{dhi1979} is modeled from a two-level
atom with a fixed energy gap. The detector works by interacting with an
external field that has been prepared in a given initial state \cite{wgu1976}.
Since the response of the detector is only based on the particle content of
the field, it can be used to study the quantum field in any spacetime without
relying on spacetime symmetry. Based on this, the detector model is widely
applied in many different situations (see the review \cite{chm2008} and
references therein). However, the field contents only for the commutative
spacetime (irrespective of the spacetime is flat or curved) were studied in
the past several years as far as we know, so in this paper, we will
investigate the response of the detector outside a noncommutative
Schwarzschild black hole \cite{nss2006}.

The study of noncommutative geometry is probably stemmed from the singularity
problem in General Relativity and it is believed that the spacetime
noncommutativity can cure the divergence in the singular points that exists in
the commutative spacetime \cite{dn2001,ss2003}. The special interest in the
noncommutativity is sparked by the prediction of string theory along with the
brane-world scenario \cite{ew1996,aad1998,sw1999}. The noncommutativity was
also introduced into the black holes since it can remove the so-called Hawking
paradox where the temperature diverges as the radius of a commutative black
hole shrinks to zero \cite{acn1987,coy2015}, and adoption of spacetime
noncommutativity leads to different black holes
\cite{nss2006,los2006,anss2007,ctz2007,ms2008,ssn2009,bcg2009} and their
associated thermodynamics \cite{bms2008,pn2009,zcy2011,zy2017}. Since the
black hole images
\cite{ka2019-1,ka2019-2,ka2019-3,ka2019-4,ka2019-5,ka2019-6,ka2022} can be
observed, it is possible to glean some information about the structure of the
black hole \cite{cjw2023}. If the noncommutativity is the fundamental property
of nature, the general black holes should carry the information about the
spacetime noncommutativity. So it is significant to study the influence of
noncommutative black holes on the physics outside the black hole.

In this paper, we numerically investigate the UDW detector coupled to a
massless scalar field in the noncommutative Schwarzschild spacetime under the
Hartle-Hawking state by extending the method of Ref. \cite{hlo2014} to the
noncommutative case. We calculate the transition rate of a static two-level
atom at a fixed position outside the black hole for some different situations,
and compare the results with those of the commutative Schwarzschild black
hole. We also use the quantum Fisher information (QFI) \cite{bc1994} to study
the possibility of extracting the information about noncommutativity from the
transition rate of the atom. In this paper, we use units with $\hbar=c=G=1$.

\section{Noncommutative black hole and Hartle-Hawking state}

In this paper, we adopt the idea assumed in Ref. \cite{nss2006} where the
spatial noncommutative effect is attributed to the modified energy--momentum
tensor as a source while the Einstein tensor is not changed, since the direct
application of noncommutative coordinates to black holes is inconvenient.
Along the line, we can change the mass of a gravitational body to include the
noncommutative effect in gravity. Thus, the smearing mass could be taken by
redefining its density through a Gaussian distribution of minimal width
$\sqrt{\theta}$ ($\theta$ is the noncommutative parameter with the dimension
length squared) instead of the Dirac delta function for the usual definition
of mass density in commutative space. With the mass density for a smearing
mass, $\rho_{\theta}(r)=\frac{M}{(4\pi\theta)^{3/2}}\exp\left(  -r^{2}%
/4\theta\right)  $, one can solve the Einstein field equation to obtain the
metric for the noncommutative black hole as%
\begin{equation}
ds^{2}=-fdt^{2}+\frac{1}{f}dr^{2}+r^{2}d\Omega^{2}, \label{nbh}%
\end{equation}
where $f\left(  r\right)  =1-\frac{4M}{r\sqrt{\pi}}\gamma(\frac{3}{2}%
,\frac{r^{2}}{4\theta})$, and the lower incomplete gamma function
$\gamma(3/2,r^{2}/4\theta)=\int_{0}^{r^{2}/4\theta}t^{1/2}e^{-t}dt$ which
approaches to $\sqrt{\pi}/2$ as $r\rightarrow\infty$. In the $\theta
\rightarrow0$ limit, the incomplete $\gamma$-function becomes the usual gamma
function and the noncommutative metric in Eq. (\ref{nbh}) becomes the
commutative Schwarzschild metric. From the condition of $f\left(  r\right)
=0$, the event horizon can be found as $r_{H}=\frac{4M}{\sqrt{\pi}}%
\gamma(3/2,r_{H}^{2}/4\theta)\equiv\frac{4M}{\sqrt{\pi}}\gamma_{H}$. The
temperature is obtained for the noncommutative metric (\ref{nbh}) as
$T_{H}=\frac{1}{4\pi}\frac{df\left(  r\right)  }{dr}|_{r=r_{H}}=\frac{1}{4\pi
r_{H}}\left(  1-\frac{r_{H}^{3}}{4\theta^{3/2}\gamma_{H}}e^{-\frac{r_{H}^{2}%
}{4\theta}}\right)  $ which decreases to zero at $M=M_{0}\simeq1.9\sqrt
{\theta}$ determined by $dM/dr_{H}=0$.

The modal solutions of the Klein-Gordon equation, $\square\Psi=\frac{1}%
{\sqrt{-g}}\partial_{\mu}(\sqrt{-g}g^{\mu\nu}\partial_{\nu}\Psi)$, in
non-commutative spacetime are expressed as $\Psi=\frac{1}{\sqrt{4\pi\omega}%
}\frac{1}{r}\rho_{\omega l}(r)Y_{\ell m}\left(  \theta,\phi\right)
e^{-i\omega t}$ where $\omega>0$, and $Y_{\ell m}$ is a spherical harmonic
function. Then, we separate the radial equation, which is obtained by using
the tortoise coordinate $r^{\ast}=\int\frac{dr}{f\left(  r\right)  }$ as
\begin{equation}
\frac{d^{2}\rho_{\omega l}(r)}{dr^{\ast2}}+\left[  \omega^{2}-V(r)\right]
\rho_{\omega l}(r)=0, \label{nre}%
\end{equation}
where $V\left(  r\right)  =f\left(  r\right)  \left[  \frac{l\left(
l+1\right)  }{r^{2}}+\frac{1}{r}\frac{df\left(  r\right)  }{dr}\right]  $.

To solve the equation, the boundary condition must be known. It is not hard to
obtain that%
\begin{align}
\rho_{\omega l}(r)  &  \sim e^{i\omega r^{\ast}},as\text{ }r^{\ast}%
\rightarrow-\infty\text{ }or\text{ }r\rightarrow r_{H},\nonumber\\
\rho_{\omega l}(r)  &  \sim e^{-i\omega r^{\ast}},as\text{ }r^{\ast
},r\rightarrow\infty,
\end{align}
which has the same form as the commutative case as given in Ref.
\cite{hlo2014}. Thus, we use the same method as in Ref. \cite{hlo2014} to
numerically solve the scalar field $\Psi$.

For the noncommutative black hole, the Hartle-Hawking state is also regular
across the both the past and future horizon, and it reduces to a thermal field
at spatial infinity with the temperature $T_{H}$ for the noncommutative black
hole. For the Hartle-Hawking state, the basic modes have the properties of
positive-frequency plane waves with respect to the horizon generators and are
expressed as \cite{cf1977,bd1984},
\begin{align}
w_{\omega lm}^{in}  &  =\frac{1}{\sqrt{2\sinh\left(  \frac{\omega}{2T_{H}%
}\right)  }}\left(  e^{\frac{\omega}{4T_{H}}}u_{\omega lm}^{in}+e^{-\frac
{\omega}{4T_{H}}}v_{\omega lm}^{in\ast}\right)  ,\nonumber\\
\bar{w}_{\omega lm}^{in}  &  =\frac{1}{\sqrt{2\sinh\left(  \frac{\omega
}{2T_{H}}\right)  }}\left(  e^{-\frac{\omega}{4T_{H}}}u_{\omega lm}^{in\ast
}+e^{\frac{\omega}{4T_{H}}}v_{\omega lm}^{in}\right)  ,\nonumber\\
w_{\omega lm}^{up}  &  =\frac{1}{\sqrt{2\sinh\left(  \frac{\omega}{2T_{H}%
}\right)  }}\left(  e^{\frac{\omega}{4T_{H}}}u_{\omega lm}^{up}+e^{-\frac
{\omega}{4T_{H}}}v_{\omega lm}^{up\ast}\right)  ,\nonumber\\
\bar{w}_{\omega lm}^{up}  &  =\frac{1}{\sqrt{2\sinh\left(  \frac{\omega
}{2T_{H}}\right)  }}\left(  e^{-\frac{\omega}{4T_{H}}}u_{\omega lm}^{up\ast
}+e^{\frac{\omega}{4T_{H}}}v_{\omega lm}^{up}\right)  , \label{modes}%
\end{align}
where \textquotedblleft in\textquotedblright\ represents the ingoing modes at
infinity, \textquotedblleft up\textquotedblright\ represents the outgoing
modes at the horizon, and they have the same meaning as that in Ref.
\cite{hlo2014} but the mass is modified by the noncommutative effect. The
function $u$ is the normalized form of the scalar field $\Psi$, and the
function $\upsilon$ is analogous to $u$ on the second exterior of the Kruskal
manifold. Thus, the quantum field under the Hartle-Hawking state can be
expanded as $\psi=\sum_{\ell=0}^{\infty}\sum_{m=-\ell}^{+\ell}\int_{0}%
^{\infty}d\omega(d_{\omega\ell m}^{up}w_{\omega\ell m}^{up}+\bar{d}%
_{\omega\ell m}^{up}\bar{w}_{\omega\ell m}^{up}+d_{\omega\ell m}^{in}%
w_{\omega\ell m}^{in}+\bar{d}_{\omega\ell m}^{in}\bar{w}_{\omega\ell m}%
^{in})+h.c.$, where the annihilation and creation operators satisfies the
following formula, $\left[  d_{\omega\ell m}^{a},d_{\omega^{\prime}%
\ell^{\prime}m^{\prime}}^{a^{\prime}\dagger}\right]  =\delta\left(
\omega-\omega^{\prime}\right)  \delta_{aa^{\prime}}\delta_{\ell\ell^{\prime}%
}\delta_{mm^{\prime}}$ and $\left[  \bar{d}_{\omega\ell m}^{a},\bar{d}%
_{\omega^{\prime}\ell^{\prime}m^{\prime}}^{a^{\prime}\dagger}\right]
=\delta\left(  \omega-\omega^{\prime}\right)  \delta_{aa^{\prime}}\delta
_{\ell\ell^{\prime}}\delta_{mm^{\prime}}$ (\textquotedblleft$a$%
\textquotedblright\ represents the \textquotedblleft in\textquotedblright\ or
\textquotedblleft up\textquotedblright\ mode).

We calculate the Wightman function of the quantum field in the Hartle-Hawking
vacuum state $|0_{H}\rangle$ as
\begin{align}
W(\tau,\tau^{\prime})  &  =\langle0_{H}|\psi(x\left(  \tau\right)
)\psi(x\left(  \tau^{\prime}\right)  )|0_{H}\rangle\nonumber\\
&  =\sum_{\ell}\int_{0}^{\infty}d\omega\frac{(2\ell+1)}{16\pi^{2}\omega
\sinh\left(  \frac{\omega}{2T_{H}}\right)  }\nonumber\\
&  \times\left(  |\Phi_{\omega\ell}^{up}(R)|^{2}+|\Phi_{\omega\ell}%
^{in}(R)|^{2}\right) \nonumber\\
&  \times\cosh\left[  \frac{\omega}{2T_{H}}-\frac{i\omega\Delta\tau}%
{\sqrt{f(R)}}\right]  , \label{wfqf}%
\end{align}
where $\Phi_{\omega\ell}(r)$ is the normalized radial function $\rho_{\omega
l}(r)$ according to the Wronskian relation \cite{hlo2014}, $R$ is the position
of the detector, $\Delta\tau=\int\frac{dt}{f\left(  r\right)  }$ is the proper
time difference along the time-like geodesic trajectory, and $\Phi_{\omega
\ell}^{up,in}(R)$ is given by solving the Eq. (\ref{nre}) and makes the normalization.

\section{Transition rate}

Consider a point-like two-level atom (detector) with the ground state
$|g\rangle$ and the excited state $|e\rangle$ and it is placed outside the
noncommutative black hole and holds still there. The atom will interact with
the field $\phi$ in the Hartle-Hawking state and evolves according to the
Hamiltonian, $\hat{H}_{int}=\lambda\chi(\tau)\mu(\tau)\phi(x(\tau))$ where
$\lambda\ll1$ is the coupling constant, $\chi(\tau)=\exp(-\tau^{2}/2\sigma
^{2})$ is a time-varying switching function, and $\mu(\tau)$ is a SU(2) ladder
operator describing the change of the atom, defined as $\mu(\tau
)=|e\rangle\langle g|e^{iE\tau}+|g\rangle\langle e|e^{-iE\tau}$ ($E$ is the
energy difference between two levels of the atom). The atom moves along a
trajectory $x(\tau)$ in the noncommutative spacetime where $\tau$ is the
proper time of the detector. In the interaction picture, the time evolution
operator is expressed as $\hat{U}=\mathcal{T}\exp\left(  -i\int d\tau\hat
{H}_{int}(\tau)\right)  $ where $\mathcal{T}$ represents the time order.
Expanding the operator to the first order, we obtain%
\begin{equation}
\hat{U}=\hat{U}^{(0)}+\hat{U}^{(1)}+\mathcal{O}(\lambda^{2}), \label{hia}%
\end{equation}
where $\hat{U}^{(0)}=\mathbb{I}$ and $\hat{U}^{(1)}=-i\int d\tau\hat{H}_{int}$.

For the initial state of the atom with the ground state $|e\rangle$ and the
field with the state $|0_{H}\rangle$, the density matrix is written as
$\hat{\rho}_{i}=|e\rangle\langle e|\otimes|0_{H}\rangle\langle0_{H}|$. Through
the evolution of unitary operators, we get the final state's density operator
\begin{align}
\hat{\rho}_{f}  &  =Tr_{\phi}[\hat{U}\hat{\rho}_{i}\hat{U}^{\dagger
}]\nonumber\\
&  =\hat{U}^{(0)}\hat{\rho}_{i}\hat{U}^{(0)\dagger}+\hat{U}^{(1)}\hat{\rho
}_{i}\hat{U}^{(1)\dagger}+\mathcal{O}(\lambda^{4})\nonumber\\
&  =|0_{H}\rangle|e\rangle\langle e|\langle0_{H}|\nonumber\\
&  +\lambda^{2}\int_{-\infty}^{\infty}d\tau\int_{-\infty}^{\infty}%
d\tau^{\prime}\chi(\tau)\chi(\tau^{\prime})\nonumber\\
&  \times e^{-iE(\tau-\tau^{\prime})}W(\tau,\tau^{\prime})|0_{H}%
\rangle|g\rangle\langle g|\langle0_{H}|+\mathcal{O}(\lambda^{4}), \label{dof}%
\end{align}
where the field state has been traced out.

\begin{figure}[ptb]
\centering
\includegraphics[width=1\columnwidth]{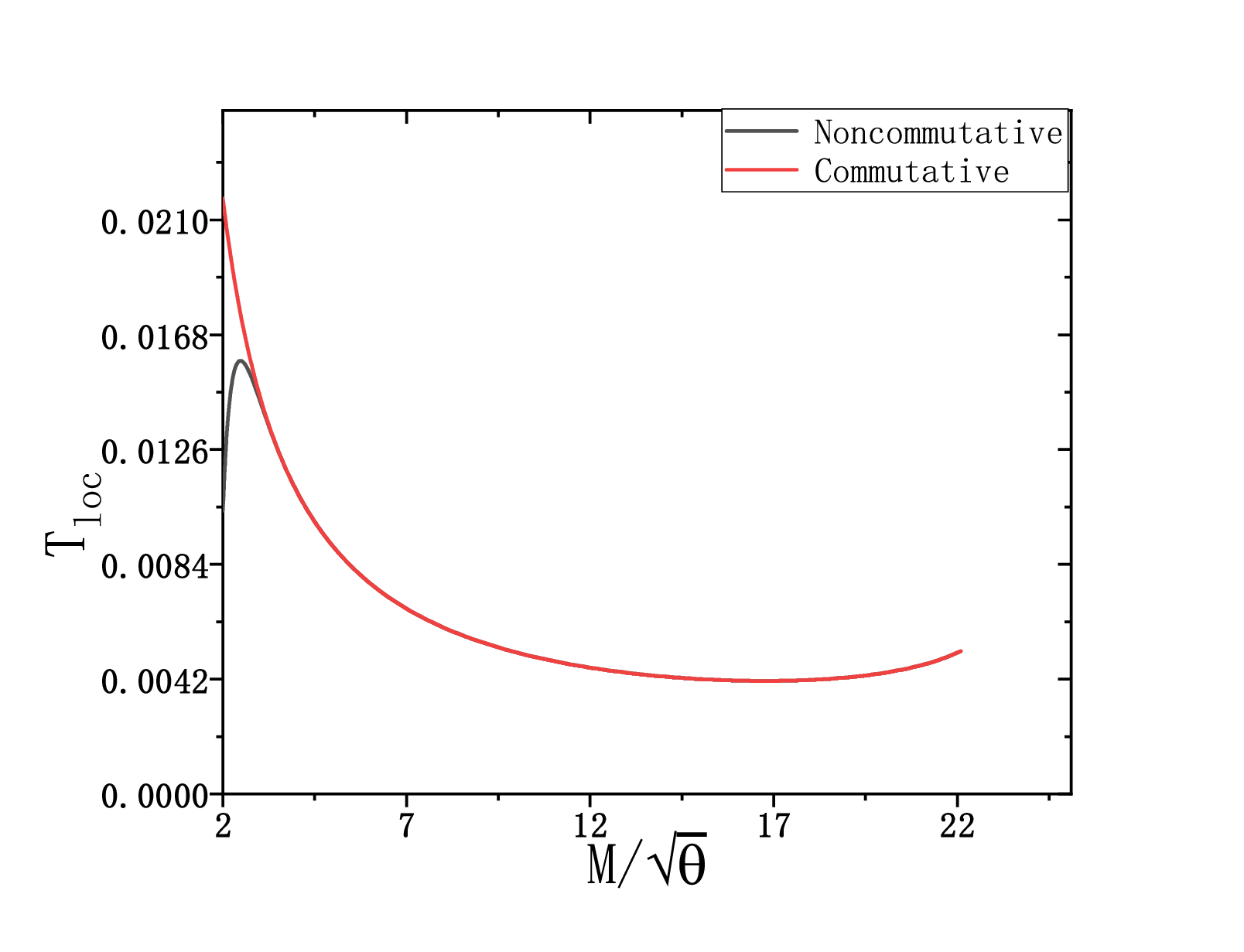}\newline%
\includegraphics[width=1\columnwidth]{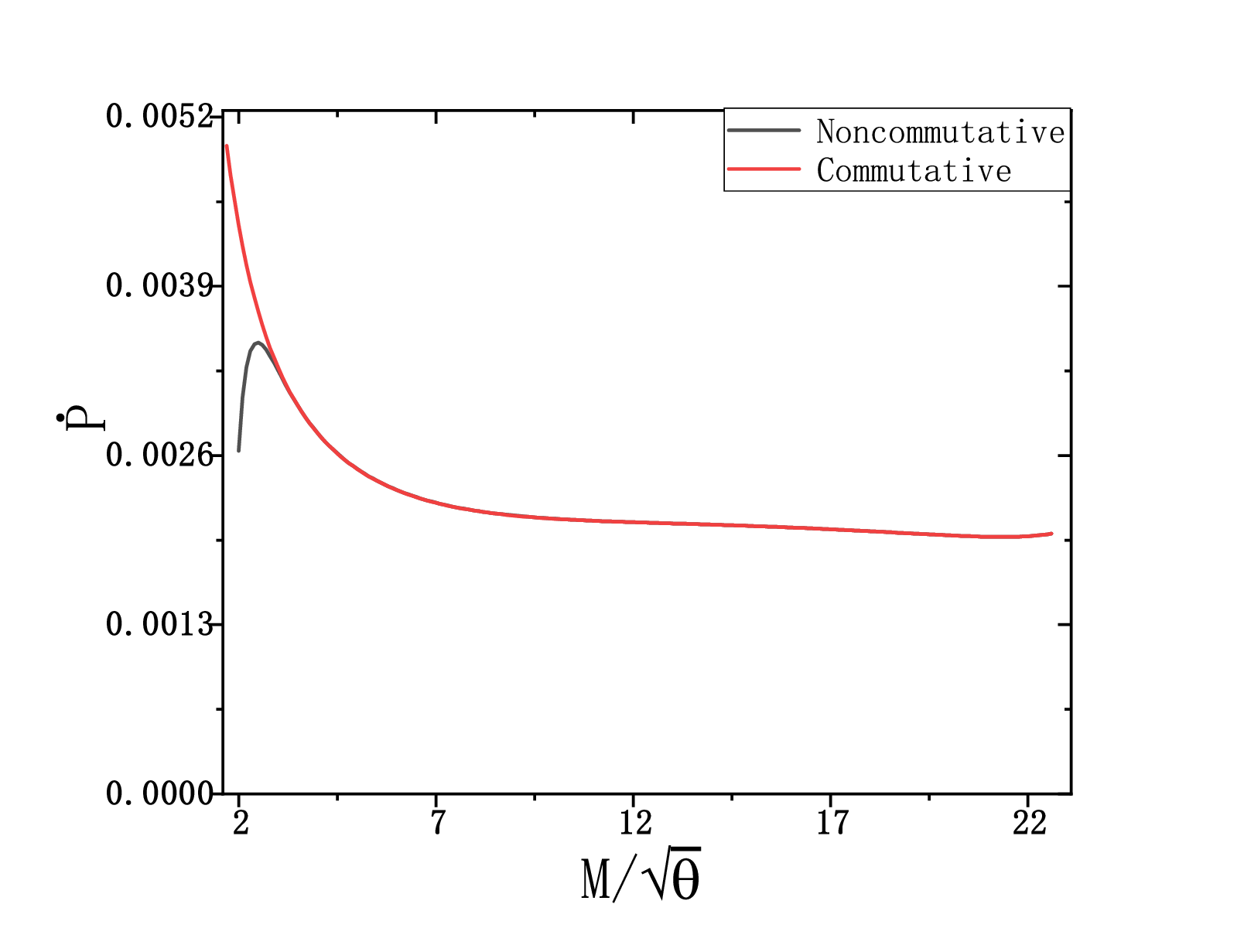}\newline\caption{ The local
temperature (upper panel) and the transition rate (lower panel) as a function
of $M$ for the case in which the detector is fixed at a position outside the
black hole. The model parameters employed are $E=0.01$ and $R=50$. }%
\label{Fig1}%
\end{figure}

The transition probability of the detector is read from the second term of the
density operator (\ref{dof}),
\begin{equation}
\mathcal{P}(e)=\lambda^{2}\int_{-\infty}^{\infty}d\tau\int_{-\infty}^{\infty
}d\tau^{\prime}\chi(\tau)\chi(\tau^{\prime})e^{-iE(\tau-\tau^{\prime})}%
W(\tau,\tau^{\prime}).\label{pdo}%
\end{equation}
Since the Wightman function satisfies the time translation invariance along
the static trajectory, one can formally drop the $\tau^{\prime}$-integral of
Eq. (\ref{pdo}) with a change of variables. Then, by first factoring out the
total effective interaction time and subsequently taking the infinite
interaction time limit $\sigma\rightarrow\infty$, the transition rate is
obtained as%
\begin{equation}
\dot{\mathcal{P}}\left(  e\right)  =\int_{-\infty}^{\infty}e^{-iEs}%
W(s)ds,\label{trd}%
\end{equation}
where $s=\tau-\tau^{\prime}$ and $\lambda=1$ for convenience.

Using the result of Eq. (\ref{wfqf}), the transition rate is given with an
explicit form as
\begin{align}
\dot{\mathcal{P}}(e)  &  =\sum_{\ell=0}^{\infty}\frac{(2\ell+1)}{4\pi}%
\sqrt{f(R)}\left(  |\Phi_{\tilde{\omega}\ell}^{up}(R)|^{2}+|\Phi
_{\tilde{\omega}\ell}^{in}(R)|^{2}\right)  \times\nonumber\\
&  [\frac{e^{-\frac{\tilde{\omega}}{2T_{H}}}\Theta(-E)}{-4\tilde{\omega}%
\sinh\left(  -\frac{\tilde{\omega}}{2T_{H}}\right)  }\nonumber\\
&  +\frac{e^{-\frac{\tilde{\omega}}{2T_{H}}}\Theta(E)}{4\tilde{\omega}%
\sinh\left(  \frac{\tilde{\omega}}{2T_{H}}\right)  }]\nonumber\\
&  =\frac{1}{8\pi E}\frac{1}{e^{E/T_{loc}}-1}\sum_{l=0}^{\infty}%
(2\ell+1)\left(  |\Phi_{\tilde{\omega}\ell}^{up}(R)|^{2}+|\Phi_{\tilde{\omega
}\ell}^{in}(R)|^{2}\right)  \label{ftr}%
\end{align}
where $\tilde{\omega}=E\sqrt{f(R)}$, and $T_{loc}$ refers to the local Hawking
temperature,
\begin{equation}
T_{loc}=\frac{T_{H}}{\sqrt{f(R)}}. \label{lht}%
\end{equation}
It is not hard to confirm that this result satisfies Kubo-Martin-Schwinger
(KMS) condition \cite{rk1957,ms1959} of thermal balance with the local
temperature $T_{loc}$.

\begin{figure}[ptb]
\centering
\includegraphics[width=1\columnwidth]{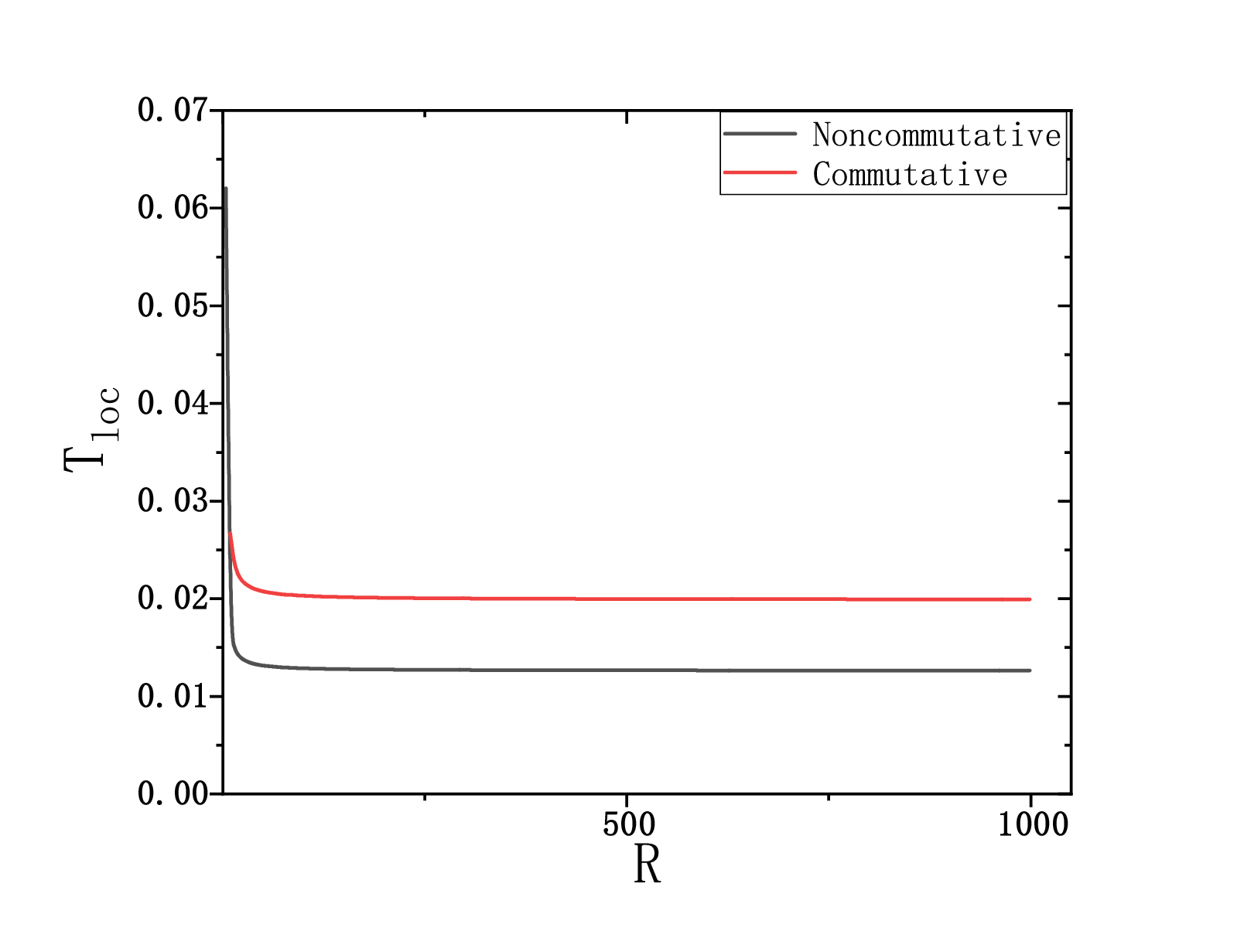}\newline%
\includegraphics[width=1\columnwidth]{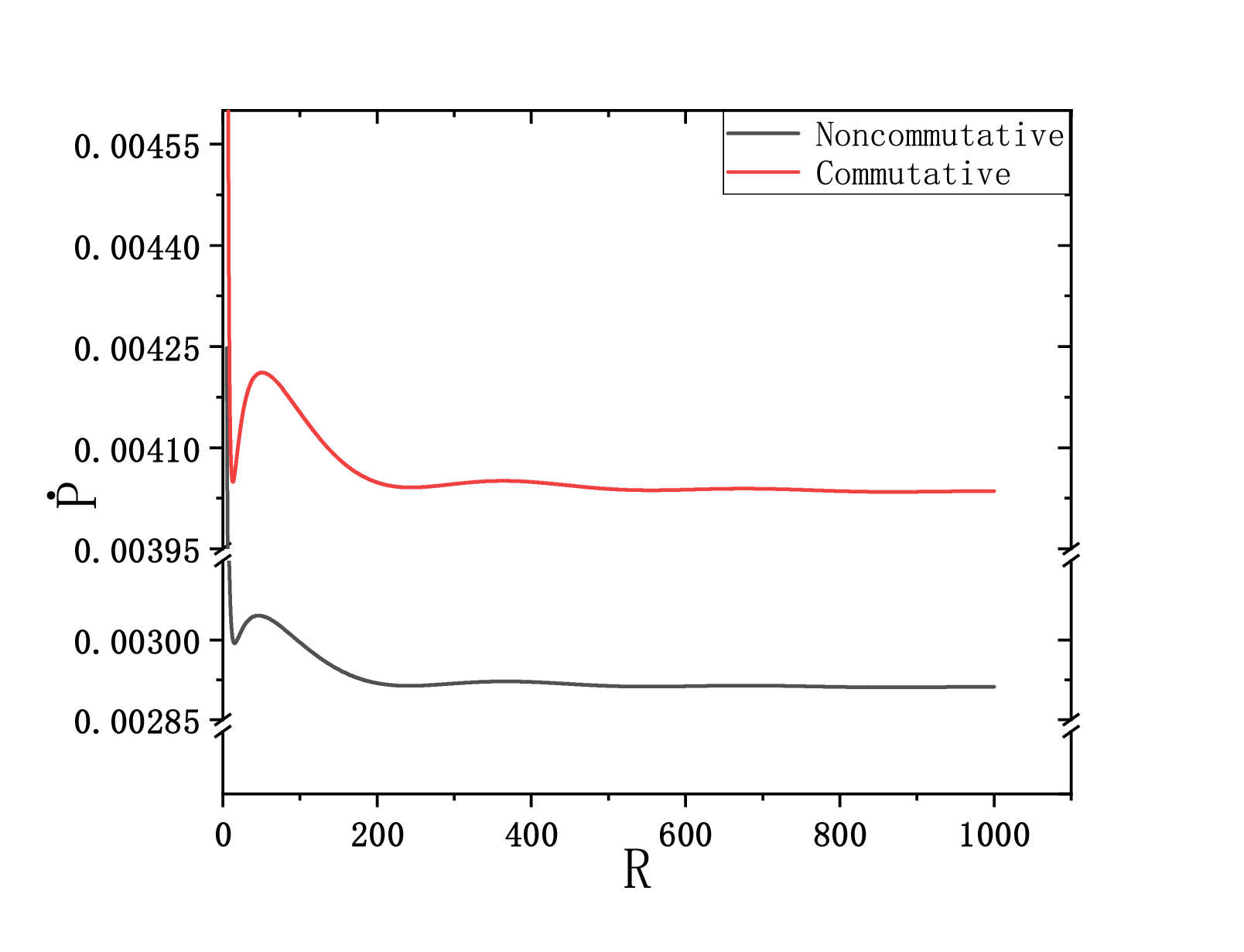}\newline\caption{ The local
temperature (upper panel) and the transition rate (lower panel) as a function
of $R$ for the case in which the mass of the black hole is fixed. The model
parameters employed are $E=0.01$ and $M/\sqrt{\theta}=2$.}%
\label{Fig2}%
\end{figure}

The upper panel of Fig. 1 presents the change of the local temperature at the
distance $R$ outside the noncommutative black hole. It is found that the local
temperature also reduces to zero when the mass approaches to the minimum value
$M_{0}$, similar to the behavior of $T_{H}$. This is interesting that the
local temperature is higher than $T_{H}$ when $M>M_{0}$, but both of them
decay to zero when $M\rightarrow M_{0}$. The lower panel of Fig. 1 presents
the transition rate for the case in which the detector is fixed at the
position outside the black hole and the mass of the black hole is changed. As
expected, the transition rate approaches to zero when $M\rightarrow M_{0}$
where the local temperature is also zero.

We also study the change of local temperature and the transition rate for the
case in which the mass of the black hole is fixed at a small value and the
distance between the detector and the horizon is changed. The upper panel of
Fig. 2 presents the results for the local temperature. It is seen that the
local temperature for the noncommutative black hole is lower than that for the
commutative black hole and both of them will diverge when the detector
approaches to the horizon since the red shift factor exists in the
denominator. The lower panel of Fig. 2 presents the results for the transition
rate. It shows that the farther the distance, the smaller the transition rate,
which is consisten with the situation as in the lower panel of Fig. 1 where it
is seen that the transition rate decreases for the noncommutative black hole
but it increases for the commutative black hole when the mass of the black
hole become small enough.

\section{Fisher information}

To analyze the possibility of measuring the noncommutativity that exists in
the black hole spacetime, we use QFI to estimate it. The QFI is a central
quantity in quantum metrology and the quantum analogue of the classical Fisher information.

\begin{figure}[ptb]
\centering
\includegraphics[width=1\columnwidth]{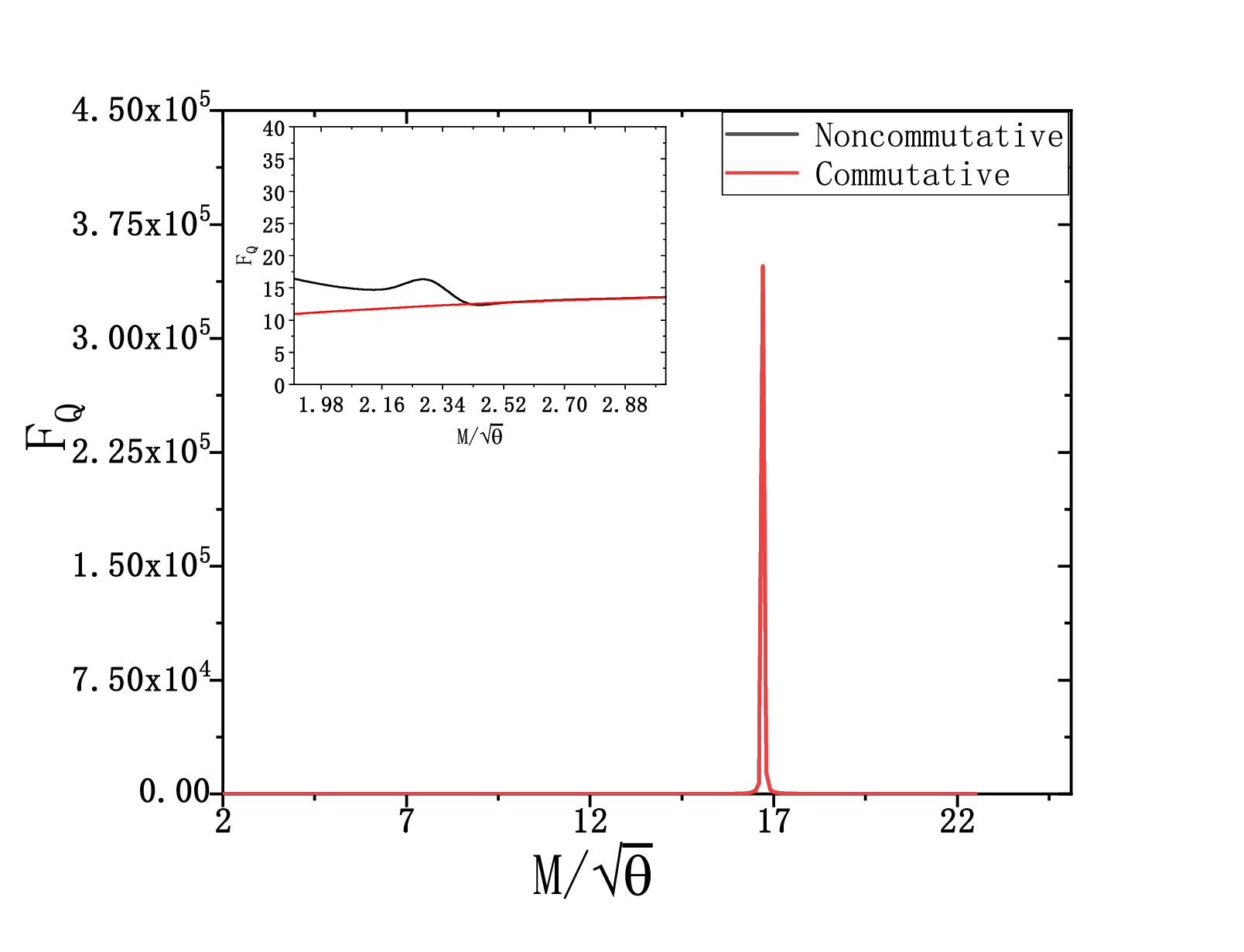} \caption{The QFI as a function
of $M$ for the case in which the detector is fixed at a position outside the
black hole. The inserted figure presents the behaviors of QFI for the case
that the mass of the black hole is small. The model parameters are the same as
in Fig. 1. }%
\label{Fig3}%
\end{figure}

According to the quantum Cram\'{e}r-Rao theorem, for a given observable
temperature $T_{loc}$, the measurement precision is determined by
\cite{bc1994,bcm1996}
\begin{equation}
Var(T_{loc})\geq\frac{1}{n\mathcal{F}_{Q}(T_{loc})}%
\end{equation}
where $Var$ represents the covariant variance, and $n$ represents the number
of repeated measurements. $\mathcal{F}_{Q}$ is just the QFI defined by
\cite{mgp2009} $\mathcal{F}_{Q}\left(  T\right)  =2\sum_{m,n}^{N}%
\frac{|\langle\psi_{m}|\partial_{T}\rho_{\omega\ell}|\psi_{n}\rangle|}%
{p_{m}+p_{n}}=\sum_{m^{\prime}}\frac{\left(  \partial_{T}p_{m^{\prime}%
}\right)  ^{2}}{p_{m^{\prime}}}+2\sum_{m\neq n}\frac{(p_{m}-p_{n})^{2}}%
{p_{m}+p_{n}}\left\vert \langle\psi_{m}|\partial_{T}\psi_{n}\rangle\right\vert
^{2}$ where $\rho_{\omega\ell}=\sum_{m=1}^{N}p_{m}|\psi_{m}\rangle\langle
\psi_{m}|$ with $p_{m}\geq0$, $\sum_{m}^{N}p_{m}=1$. For our consideration,
the density matrix is $\rho_{\omega\ell}=\left(
\begin{array}
[c]{cc}%
1-\dot{\mathcal{P}}(e) & 0\\
0 & \dot{\mathcal{P}}(e)
\end{array}
\right)  +\mathcal{O}(\lambda^{4})$. Therefore, we obtain%
\begin{equation}
\mathcal{F}_{Q}(T)=\frac{\left(  \partial_{T_{loc}}(1-\dot{\mathcal{P}%
})\right)  ^{2}}{\dot{\mathcal{P}}}+\frac{\left(  \partial_{T_{loc}}%
\dot{\mathcal{P}}\right)  ^{2}}{\dot{\mathcal{P}}}=2\frac{\left(
\partial_{T_{loc}}\dot{\mathcal{P}}\right)  ^{2}}{\dot{\mathcal{P}}}%
\end{equation}

\begin{figure}[ptb]
\centering
\includegraphics[width=1\columnwidth]{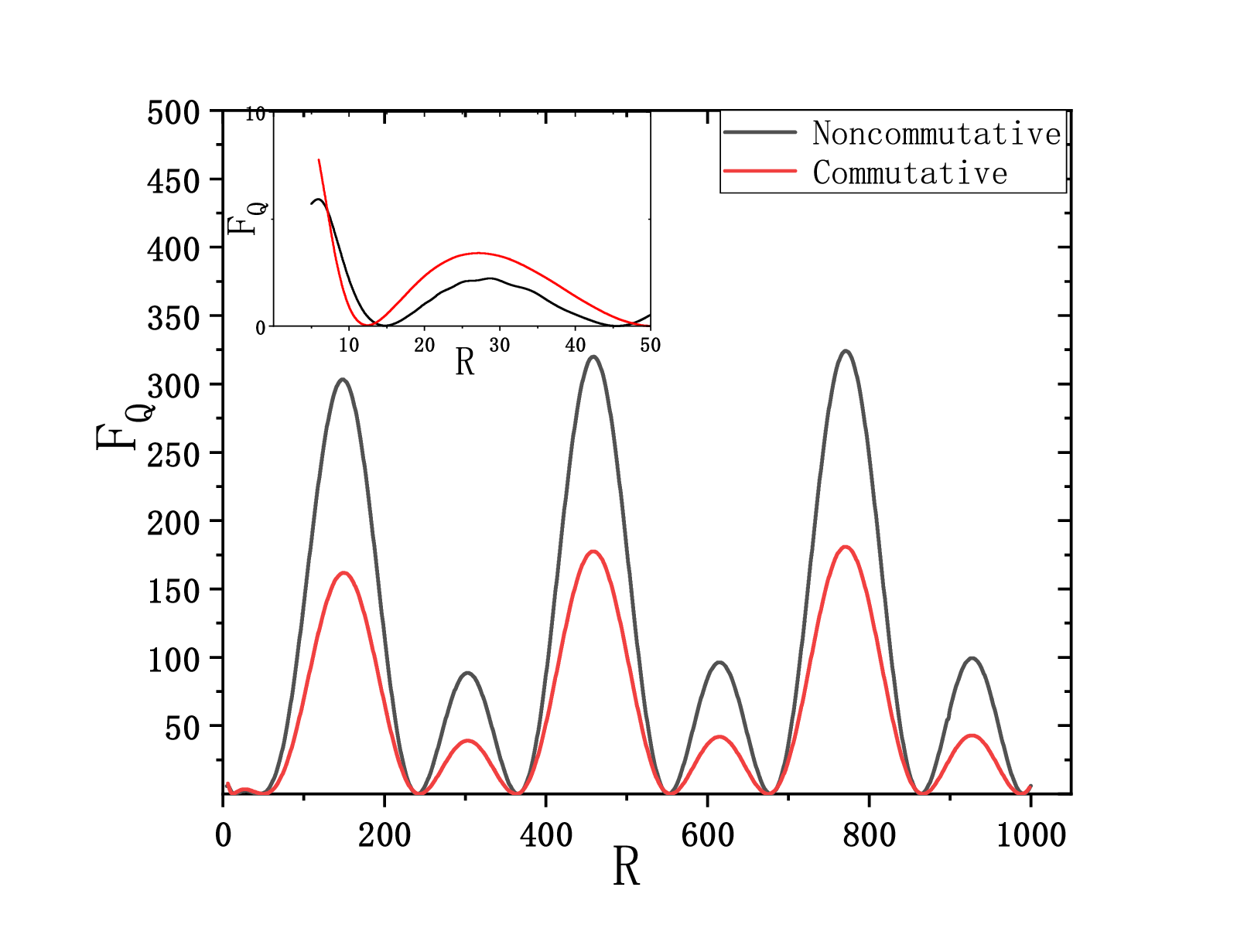} \caption{The QFI as a function
of $R$ for the case in which the detector is fixed at a position outside the
black hole. The inserted figure presents the behaviors of QFI for the case
that the distance between the detector and the horizon is short. The model
parameters are the same as in Fig. 2. }%
\label{Fig4}%
\end{figure}

Figure 3 presents the change of QFI with the mass of the black hole for a
static detector fixed at a position outside the black hole, as that in Fig. 1.
It can be observed that the QFI is nearly the same for both commutative and
noncommutative black holes. However, a difference emerges after the mass
decreases, where the QFI for the noncommutative black hole surpasses that for
the commutative black hole due to the influence of spatial noncommutativity.
In particular, there is a divergent peak at $M_{d}\simeq16.67$. This can be
obtained by rewriting the QFI as $\mathcal{F}_{Q}(T)=\frac{2}{\dot
{\mathcal{P}}}\left(  \frac{\partial_{M}\dot{\mathcal{P}}}{\partial_{M}%
T_{loc}}\right)  ^{2}$ and the partial derivative $\partial_{M}T_{loc}$ is
zero at the mass $M_{d}$. It is surprising that the zero points of
$\partial_{M}T_{loc}$ are the same for the commutative and noncommutative
black holes,\ which has been confirmed by taking the numerical value to the
forty units after the decimal point.

Figure 4 gives the change of QFI with the distance between the detector and
the horizon for a black hole with a fixed small mass. It is interesting to
find that the QFI present a fluctuating behavior, and the amplitude of
fluctuation for the noncommutative black hole is larger than that for the
commutative black hole. The reason for the fluctuation is that the transition
rate presents a fluctuating behavior when the detector is placed at a far away
location, as in the lower panel of Fig. 2. But the amplitude of the
fluctuation is very small for the transition rate, so it is not easy to be
seen. The QFI enlarges the fluctuation which is helpful for the possible observation.

\section{Conclusion}

In this paper, we investigate the behavior of UDW detectors under the
background of noncommutative black holes. We study two cases: one is that the
position of the detector is fixed and the mass of the black hole is changed,
in which the local Hawking temperature and the transition rate of the detector
present the different phenomena when the mass becomes small enough; the other
one is that the position of the detector is changed for a fixed small mass of
the black hole, in which the local Hawking temperature and the transition rate
of the detector present the similar change but with different numerical
values. We also calculate the QFI for these two cases. A novel and interesting
phenomenon is found that the noncommutativity can enhance the fluctuation of
the QFI with the distance between the detector and the horizon for a fixed
small mass of the black hole. In particular, the fluctuation is still strong
enough even for a far enough distance, which is helpful for the possible
future observation.

\section{Acknowledgments}

We are grateful to the anonymous referee for his/her critical comments and
helpful advice. This work is supported by National Natural Science Foundation
of China (NSFC) with Grant No. 12375057, and the Fundamental Research Funds
for the Central Universities, China University of Geosciences (Wuhan) with No. G1323523064.

\end{document}